\documentclass[%
 reprint,
 amsmath,amssymb,
 aps,
]{revtex4-2}

\usepackage{graphicx}% Include figure files
\usepackage{dcolumn}% Align table columns on decimal point
\usepackage{bm}% bold math
\usepackage{lipsum, babel}
\usepackage{soul}
\usepackage{comment}
\usepackage[svgnames]{xcolor}
\usepackage{xstring}
\newcommand{\EatOneArg}[1]{}

\newcommand{\addMS}[1]{\textcolor{black}{#1}}
\newcommand{\addOB}[1]{\textcolor{black}{#1}}
\newcommand{\addOBB}[1]{\textcolor{black}{#1}}
\newcommand{\addMSC}[1]{\textcolor{black}{#1}}
\newcommand{\addMSS}[1]{\textcolor{black}{#1}}
\usepackage{graphicx}% Include figure files
\usepackage{dcolumn}% Align table columns on decimal point
\usepackage{bm}% bold math
\usepackage[hidelinks]{hyperref}% add hypertext capabilities

\usepackage{xcolor}

\newcommand{\diff}{\mathrm{d}}

\newcommand{\bx}{\mathbf{x}}
\newcommand{\bk}{\mathbf{k}}

\newcommand{\oma}{\omega}

\newcommand{\cg}{c_g}
\newcommand{\half}{\frac{1}{2}}
\newcommand{\Za}{Z_\alpha}
\newcommand{\Zb}{Z_\beta}

\newcommand{\ea}{e_\alpha}
\newcommand{\eb}{e_\beta}
\newcommand{\ec}{e_\gamma}

\renewcommand{\oma}{\omega_\alpha}
\newcommand{\omb}{\omega_\beta}
\newcommand{\omg}{\omega_\gamma}

\renewcommand{\Re}{\operatorname{Re}}

\begin{document}

\setlength{\parindent}{0cm}

%lyx shortcuts
\global\long\def\dyad#1{\underline{\underline{\boldsymbol{#1}}}}%
\global\long\def\ubar#1{\underbar{\ensuremath{\boldsymbol{#1}}}}%
\global\long\def\integer{\mathbb{Z}}%
\global\long\def\natural{\mathbb{N}}%
\global\long\def\real#1{\mathbb{R}^{#1}}%
\global\long\def\complex#1{\mathbb{C}^{#1}}%
\global\long\def\defined{\triangleq}%
\global\long\def\trace{\text{trace}}%
\global\long\def\del{\nabla}%
\global\long\def\cross{\times}%
\global\long\def\diff#1#2{\frac{\partial#1}{\partial#2}}%
\global\long\def\Diff#1#2{\frac{d#1}{d#2}}%
\global\long\def\bra#1{\left\langle #1\right|}%
\global\long\def\ket#1{\left|#1\right\rangle }%
\global\long\def\braket#1#2{\left\langle #1|#2\right\rangle }%
\global\long\def\ketbra#1#2{\left|#1\right\rangle \left\langle #2\right|}%
\global\long\def\identity{\mathbf{1}}%
\global\long\def\paulix{\begin{pmatrix}  &  1\\
 1 
\end{pmatrix}}%
\global\long\def\pauliy{\begin{pmatrix}  &  -i\\
 i 
\end{pmatrix}}%
\global\long\def\pauliz{\begin{pmatrix}1\\
  &  -1 
\end{pmatrix}}%
\global\long\def\sinc{\mbox{sinc}}%
\global\long\def\four{\mathcal{F}}%
\global\long\def\dag{^{\dagger}}%
\global\long\def\norm#1{\left\Vert #1\right\Vert }%
\global\long\def\hamil{\mathcal{H}}%
\global\long\def\tens{\otimes}%
\global\long\def\ord#1{\mathcal{O}\left(#1\right)}%
\global\long\def\undercom#1#2{\underset{_{#2}}{\underbrace{#1}}}%
 
\global\long\def\conv#1#2{\underset{_{#1\rightarrow#2}}{\longrightarrow}}%
\global\long\def\tg{^{\prime}}%
\global\long\def\ttg{^{\prime\prime}}%
\global\long\def\clop#1{\left[#1\right)}%
\global\long\def\opcl#1{\left(#1\right]}%
\global\long\def\broket#1#2#3{\bra{#1}#2\ket{#3}}%
\global\long\def\div{\del\cdot}%
\global\long\def\rot{\del\cross}%
\global\long\def\up{\uparrow}%
\global\long\def\down{\downarrow}%
\global\long\def\Tr{\mbox{Tr}}%

\global\long\def\per{\mbox{}}%
\global\long\def\pd{\mbox{}}%
\global\long\def\p{\mbox{}}%
\global\long\def\ad{\mbox{}}%
\global\long\def\a{\mbox{}}%
\global\long\def\la{\mbox{\ensuremath{\mathcal{L}}}}%
\global\long\def\cm{\mathcal{M}}%
\global\long\def\cg{\mbox{\ensuremath{\mathcal{G}}}}%
%end lyx shortcuts

\preprint{APS/123-QED}

\title{Turbulent spectrum of 2D internal gravity waves}

\author{Michal Shavit, Oliver B\"uhler 
and Jalal Shatah}

\affiliation{%
 Courant Institute of Mathematical Sciences, New York University, NY 10012, USA.
}%


\begin{abstract}
%needs to be < 600 characters 
We find the turbulent energy spectrum of weakly interacting 2D internal gravity waves \addOB{using the full, non-hydrostatic dispersion relation.}  This spectrum is an exact solution of a \addOB{regularized kinetic equation, from which the zero-frequency shear modes have been excised by a careful limiting process.  This is a new method in wave kinetic theory.}  The turbulent spectrum agrees with the 2D oceanic Garrett--Munk spectrum for frequencies large compared to the Coriolis frequency and vertical scales small compared to the depth of the ocean. \addMS{We show that this turbulent spectrum  is the unique  \addOBB{power law solution to the steady kinetic equation} with a non-zero radial flux}.  Our solution provides an interesting insight into a turbulent energy cascade in an anisotropic system---like isotropic turbulence it is self-similar in scale, but its angular part is peaked along the curve of vanishing frequency and is \addMS{self-similar in frequency}. 
\end{abstract}

\maketitle

%\tableofcontents

%\section{\label{sec:level1}Introduction}
\textbf{\textit{Introduction}} 
%In 1966, Hasselmann laid the groundwork by introducing perturbative equations for weakly interacting geophysical waves \cite{hasselmann1966feynman}. Since then, 
Despite a wealth of oceanic measurements and continuous theoretical advancements \cite{mccomas1977resonant, muller1986nonlinear, caillol2000kinetic,lvov2001hamiltonian}, a theoretical derivation of the energy spectrum for internal gravity waves in the ocean (the renowned empirical Garrett-Munk~(GM) spectrum \cite{garrett1979internal,munk1981internal}) remains an open problem. While observations emphasize the central role dispersive internal gravity waves play in natural processes like the ocean's climate cycle \cite{whalen2020internal}, theoretically and experimentally decoupling these waves from the evolution of slow modes, degrees of freedom with vanishing frequency, proves difficult. Slow modes, such as shear and domain modes in 2D and 3D and vortical modes in 3D, are non-linearly interacting with the dispersive waves, and affect a large part of the energy spectrum \cite{lanchon2023internal,smith2001numerical,smith2002generation}.   A promising avenue lies in the kinetic approach \cite{lvov2001hamiltonian, caillol_kinetic_2000,labarre2024internal, dematteis2022}; however, the nonlinear fluid equations in both 2D and 3D yield  an anisotropic, non-canonical Hamiltonian system, which turns its kinetic description into a non-trivial problem since the classical wave turbulence methods are almost irrelevant. \addOB{To make headway on this difficult problem previous kinetic studies were restricted to low-frequency hydrostatic internal waves (which obey a much simpler, monomial dispersion relation), but this comes at the price of gross inaccuracy when higher-frequency waves are involved.}  
Compared to 3D, the 2D problem holds a few advantages: it has no vortical modes, it is cheaper for direct numerical simulation and its  recently derived \cite{shavit2024sign} kinetic equation takes a particular simple form due to the existence of a second quadratic invariant.  The relevance of the 2D internal gravity wave description extends to both experimental settings, such as long water tanks, and natural phenomena like internal tides around isolated 1D topographical features like the Hawaiian ridge \cite{LSY03}. 
\begin{figure}[h] \label{fig:2dspec}
\includegraphics[scale=0.4]{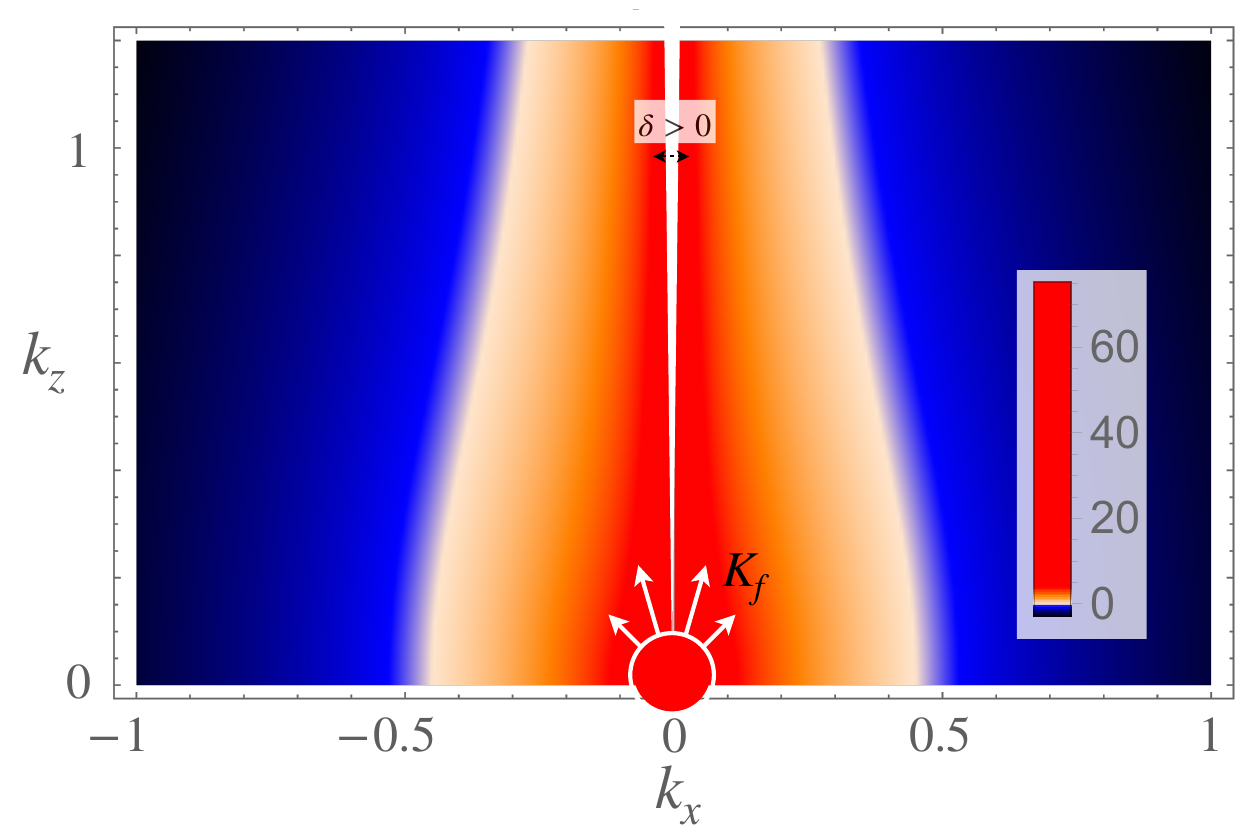}% Here is how to import EPS art
\caption{\label{fig:epsart} Turbulent energy spectrum $e_k=C_0K^{-3}(\omega_k/N)^{-2}=C_0K^{-3}\cos\theta_k^{-2}$ on the regularized $(k_x,k_z)$ plane, outside %an isotropic 
an angular source centered at radius $K_f$ and emitting a radial energy flux.}%$\Pi^r(\theta_k)=\Pi_0 \sin (\theta_k) \hat{k}/ K_f$. }
\end{figure}
In our previous work, \cite{shavit2024sign}, we derived the isotropic, angle-averaged part of the turbulent energy spectrum of 2D internal gravity waves and here we derive the angular part. \addOBB{However, in a direct approach the kinetic equation diverges around the zero-frequency shear modes in the system.   This, we claim, indicates a meaningful limit on the applicability of the kinetic description due to the existence of slow modes in the original dynamical equations. } To deal with this problem we first regularize the kinetic equation by removing all modes below some low-frequency cut-off, then we find a turbulent steady spectrum for the regularized equation, and finally we let the cut-off frequency go to zero.  This is a new method in kinetic theory, which promises to be of wider use for anisotropic dispersive wave systems in 2D or 3D, and to waves in plasma.  
\addOB{Our turbulent solution is the unique scale-invariant solution with a non-zero angular-averaged radial flux and it agrees with the 2D GM spectrum for ocean internal waves at high frequency and small scale, which is a first for a theoretically derived spectrum.}

\textbf{\textit{Governing equations}} The two-dimensional Boussinesq equations restricted to a vertical $xz$-plane can be written as 
\begin{align}\label{eq:2D Boussinesq}
       \Delta\psi_{t}+\left\{ \psi,\Delta\psi\right\} &=-N^{2}\eta_{x}\\ \eta_{t}+\left\{ \psi,\eta\right\} &=\psi_x.\nonumber
\end{align}
Here $z$ is the vertical and $x$ is the  horizontal coordinate with corresponding velocities $w$ and $u$, 
$\psi$ is a stream function such that $(\psi_x,\psi_z)=(w,-u)$ and $-\Delta \psi$ is the vorticity, $\eta$ is the vertical displacement, $N$ the constant buoyancy frequency and $\left\{ g,f\right\}=\partial_xg\partial_zf-\partial_zg\partial_xf$. The vertical buoyancy force $b=-N^2\eta$ opposes vertical displacements and derives from a consideration of potential energy  in the presence of gravity and non-uniform density. \eqref{eq:2D Boussinesq} has two quadratic invariants: the total energy $E=\int d\mathbf{x} (-\psi\Delta \psi+N^2\eta^2)$ and the pseudomomentum $P=\int d\mathbf{x} \,\eta \Delta \psi$. 
We consider a periodic domain $\mathbf{x}\in \left[0,L\right]^2$ and expand the fields $(\psi,\eta) = \sum_\alpha Z_{\alpha}(t)g_{\alpha}(\bx)$ in terms of linear wave modes, where $Z_\alpha(t)$ are complex scalar wave amplitudes and the $g_\alpha(\bx)$ are eigenvector functions for the linear part of (\ref{eq:2D Boussinesq}),
\begin{align}
-i\omega_\alpha \begin{pmatrix}-\Delta & 0\\
0 & N^{2}
\end{pmatrix} g_{\alpha} &= {N^2}\begin{pmatrix}0 & 1\\
1 & 0
\end{pmatrix}\partial_x g_{\alpha}.\label{eq:lin eigen}
\end{align}
The multi-index $\alpha=(\sigma,\mathbf{k})$ combines branch choice $\sigma =\pm 1$ and wave number $\bk\in (2\pi\mathbb{Z}/L)^2$. In polar coordinates, $\bk=K(\cos\theta_k,\sin\theta_k)$, the dispersion relation is
\begin{equation}\label{eq:omega}
\omega_{\alpha}=\sigma N \cos\theta_k. 
\end{equation}
The choice $\sigma=\pm1$ corresponds to horizontally right-going or left-going waves and $\theta_k=\pm \pi/2$, where $\omega_k=0$,  characterizes the  shear modes. \addOBB{Amplitudes evolve as 
\begin{equation}
\dot Z_\alpha+i\omega_\alpha Z_\alpha=\half\sum\limits_{\beta,\gamma} V^{\beta \gamma}_\alpha Z_\beta Z_\gamma,
\end{equation}
summing over wavenumber triads $\mathbf{k}_\alpha+\mathbf{k}_\beta+\mathbf{k}_\gamma=0$.  The interaction coefficients \cite{Joubaud12,ripa,shavit2024sign} can be written in the symmetrized form
\begin{equation}
V_\alpha^{\beta\gamma}=\frac{\mathbf{k}_{\beta}\cross\mathbf{k}_{\gamma}}{K_{\alpha}K_{\beta}K_{\gamma}}\frac{N^2}{2\sqrt{8}}\left(s_{\alpha}^{x}+s_{\beta}^{x}+s_{\gamma}^{x}\right)\left(s_\beta^x-s_{\gamma}^{x}\right).%\nonumber
\end{equation}} 
Here $s_\alpha^{x}=(k_x/\omega)_\alpha$ is the horizontal slowness.  The modal expansion diagonalizes the energy $E  = \sum _\alpha E_\alpha= \sum _\alpha Z_{\alpha}Z_{\alpha}^{*}$ and yields 
 \begin{equation}
     \label{eq: energy cont}
     \dot{E_\alpha}= \sum_{\beta,\gamma}V_\alpha^{\beta \gamma}\Re \left(Z^*_\alpha Z^*_\beta Z^*_\gamma\right).
 \end{equation} 
The wave expansion diagonalizes the pseudo-momentum as well: $P=\sum_{\alpha}  s_\alpha^{x} E_\alpha = \sum_{\alpha} \sigma_\alpha N^{-1}K E_\alpha$, so horizontally right-going waves carry positive pseudo-momentum. \addOBB{So far frequency resonance was not assumed.}  Now, the kinetic equation that describes the slow evolution of the averaged energy density $e_\alpha=\left\langle E_{\alpha}\right\rangle $, where the brackets denote averaging over an initial Gaussian  statistical ensemble,
\begin{equation}\label{eq:cor}
\left\langle \Zb^*(0)\Za(0)\right\rangle  = \delta_{\alpha\beta} e_\alpha(0),
\end{equation}
was derived in \cite{shavit2024sign} and is given by 
\begin{align}
      \label{eq:kin sim}
  \dot \ea&=St_\alpha (e_\alpha)\\ \nonumber
  &={\pi}\!\!
\int\limits_{\omega_{\alpha\beta\gamma}}\!\!\oma\,\Gamma_{\alpha\beta\gamma}^2 (\oma\eb\ec + \omb\ea\ec +
    \omg\ea\eb).%\delta(\omega_{\alpha \beta\gamma })
\end{align}
To derive the kinetic equation, which is a non-trivial closure of \eqref{eq: energy cont}, the joint kinetic limits of big box and long nonlinear times, $L\rightarrow\infty$ and $t\omega\rightarrow\infty$, are taken. So the discrete sum in \eqref{eq: energy cont}  is replaced by an integral over the resonant manifold:
\begin{equation}\label{eq:resonant manifold}
    \int\limits_{\omega_{\alpha\beta\gamma}} \!\!\!\!\!= \!\int d\beta d\gamma \,\delta(\omega_\alpha+\omega_\beta+\omega_\gamma)\delta\left(\mathbf{k}_{\alpha}+\mathbf{k}_{\beta}+\mathbf{k}_{\gamma}\right),
    \end{equation}
where $\int\!\! d\alpha=\sum_{\sigma=\pm 1} \int\!\!d\mathbf{k}$.  The  interaction coefficients are proportional to the frequencies, i.e., $V^{\beta\gamma}_\alpha/\omega_\alpha=V_\beta^{\alpha\gamma}/\omega_\beta=V_\gamma^{\alpha\beta}/\omega_\gamma=\Gamma_{\alpha\beta\gamma}$, which allows the relatively simple \eqref{eq:kin sim}. 
%\begin{equation}\label{eq:gamma}
%   \Gamma_{\alpha\beta\gamma}= \frac{1}{\sqrt{8}}\sum_{\chi=\alpha,\beta,\gamma}\!\!\!\!\sin\theta_{\chi}\!\!\!\sum_{\chi=\alpha,\beta,\gamma}\!\!\!\sigma_{\chi}K_{\chi},
%\end{equation}
In the vicinity of slow, zero-frequency modes the kinetic equation must be interpreted carefully. In the discrete sum on the lattice, \eqref{eq: energy cont}, shear modes are well separated from waves with non zero frequency, however as $L\rightarrow \infty$ the collision integral \eqref{eq:kin sim} includes integration arbitrarily close to slow modes. Slow modes cannot be created by resonant interactions, but as $\omega_\alpha\rightarrow0$ the kinetic equation needs to include off-diagonal correlators apart from \eqref{eq:cor}, such as $\left\langle {Z_\alpha^2}\right\rangle$, which would yield a very complicated description. Such correlators oscillate with frequency $\sim 2\omega_\alpha$ and over long times can be neglected as long as $\omega_\alpha\!>\!\epsilon\!>\!0$ \cite{shavit2024sign}. From that perspective, anisotropic systems with dispersion relation that vanish on a curve rather than on a point, pose a special problem for kinetic description.  This did not affect the derivation in \cite{shavit2024sign} of the homogeneous part of the energy spectrum, which depends solely on the homogeneity degree of the interaction coefficients and frequency. However, as we continue towards the derivation of the angular dependence of the energy spectrum we anticipate trouble along the $k_z$ axis, where $\omega_k=0$. To overcome this obstacle, we consider a regularized kinetic equation on the $(k_x,k_z)$ plane with a small \addMS{angular sector of width $2\delta$ around $\theta_k=\pi/2$ (corresponding to $k_z=0$) removed}, and then investigate the limit  as $\delta\rightarrow0$ numerically;  see Figure 1. To obtain our results we numerically integrate the collision integral in Mathematica (our notebook is attached to the supplementary material).

%The problem of assigning a meaningful flux to non isotropic kinetic equations has been addressed in the past \cite{dematteis2022,yokoyama2021energy}. We argue that this process is equivalent to choosing boundary conditions for the steady kinetic equation and in 2D turbulence existing between energy injection and dissipation which are separated in scale the most natural gauge fixing is the simple one that eliminates the angular component of the flux.

\textbf{\textit{Turbulent % Boundary conditions and 
steady solutions of the kinetic equation}} 
The frequency %\eqref{eq:omega} 
and the interaction coefficients are \addOB{homogeneous functions such that} $\omega_{\left(\sigma,\lambda\mathbf{k}\right)}=\omega_{\left(\sigma,\mathbf{k}\right)}$ and $V\left(\lambda\mathbf{k}_{\alpha},\lambda\mathbf{k}_{\beta},\lambda\mathbf{k}_{\gamma}\right)=\lambda V\left(\mathbf{k}_{\alpha},\mathbf{k}_{\beta},\mathbf{k}_{\gamma}\right)$ for any $\lambda>0$.
Assuming the steady spectrum $e_\alpha$ is \addOB{homogeneous as well implies } $e_{\alpha}=C_0 K^{-w} e_{\alpha}^{\theta}\left(\theta_{k}\right)$,
where $C_0>0$ is a Kolmogorov constant, \addOB{$w$ is the homogeneity degree, and $e_{\alpha}^{\theta}\left(\theta_{k}\right)$ captures the angular dependence of the spectrum \cite{shavit2024sign}.} We restrict here to the case that energy is symmetrically distributed between horizontally left-going and right-going waves, $e_{(-,\mathbf{k})}=e_{(+,\mathbf{k})}$ and then the pseudo-momentum is zero on average $\overline{P}=0$, so that $e_\alpha^\theta=e_\theta$. 
\addOB{It proves convenient to search for a solution in a form motivated by $\omega_k^2/ N^2=\cos^2\theta_k$, namely
\begin{align} \label{eq:turbspec}
e_{\theta} = (\cos^2 \theta_k) ^{f_w(\theta_k)},
\end{align}  
which also ensures $e_\alpha\geq0$. If $f_w$ depends on $\theta_k$ then \eqref{eq:turbspec} is a general \textit{Ansatz}, but we will eventually focus on a constant exponent $f_w$.}  
The homogeneous spectrum turns the collision integral (the RHS of the kinetic equation \eqref{eq:kin sim}) to a homogeneous function and hence separable in angle and wave number amplitude
 \begin{equation}\label{eq:homcoll}
  St_\alpha\equiv K^{2w_0-2w-2}C_0^2St_{\theta}\left(w,f_w, \theta_k \right).
 \end{equation}
Here $St_{\theta}$ is the angular part of the collision integral, its explicit form is found by parametrization of the resonant manifold and is presented in the supplementary material. \addMS{The previously found turbulent homogeneity degree $w_0=w_V+d-w_\omega/2$ is the sum of the homogeneity degree of the interaction coefficient $w_V=1$, dimension $d=2$ and minus half the homogeneity degree of the frequency $w_\omega=0$ \cite{shavit2024sign}.}  \addOB{Notably, for internal waves \cite{caillol_kinetic_2000}  already pointed out that the value $w_0=3$ followed from dimensional analysis based  on the units of $N$ and $\Pi_0$.}  The separability of the collision integral implies that \addOB{$w=w_0$} is the unique \addOBB{steady power law} solution carrying a non-zero radial flux. \addOBB{To show this we adapt standard arguments from isotropic turbulence \cite{ZLF}.  By energy conservation the integral of $St_\alpha$ over all modes vanishes, which under left--right symmetry allows defining}
\begin{equation}\label{eq:Pi0}
    \Pi_0(K) =- \int\limits_{|\mathbf{k}_\alpha|\leq K} St_\alpha \mathrm{d}\mathbf{k}  = + \int\limits_{|\mathbf{k}_\alpha|\geq K} St_\alpha \mathrm{d}\mathbf{k}
\end{equation}
\addOBB{as the radial flux of energy across a circle in wavenumber space with radius $K>0$.} \addMS{The term 'flux' suggests local energy transfer, but \eqref{eq:Pi0} does not require this, so 'energy exchange rate' might be a more appropriate term. We retain 'flux' based on  previous literature.}   \addOBB{Combining \eqref{eq:Pi0} with  \eqref{eq:homcoll} (and assuming $w>w_0$) an area integration over $|\mathbf{k}_\alpha|\geq K$ yields}
\begin{equation}\label{eq:fluxang} 
 \Pi_0(K)=-\frac{K^{2\left(w_{0}-w\right)}}{2(w_0-w)} C_0^2\int_0^{2\pi} St_{\theta}\,\mathrm{d}\theta.
\end{equation}
\addOBB{This shows that steady power law solutions of the kinetic equation have zero radial flux unless $w=w_0$, in which case a nonzero $K$-independent value of $\Pi_0$ can be extracted from \eqref{eq:fluxang} in the limit  $w\to w_0$.}

 \textbf{\textit{Obtaining the turbulent solution on the regularized plane.}}
 While the radial part of the energy spectrum is insensitive to the validity range of the kinetic equation as the frequency $\omega_k\rightarrow 0$,  the angular part is more delicate. To obtain the angular part of turbulent energy spectrum we first regularize the angular part of the collision integral by removing an angular sector of width $2\delta$ and then consider the limit $\delta\to0$.   This means that the angular integration over the incoming angles on the resonant manifold, \eqref{eq:resonant manifold}, is is restricted to $\theta_{\beta},\theta_{\gamma}\in\left[-\pi/2+\delta,\pi/2-\delta\right]$. In terms of frequencies, this constrains $\cos\theta_\beta,\cos\theta_\gamma \in\left[-1,-\epsilon\right]\cup\left[\epsilon,1\right]$, for small positive $\epsilon$, $\epsilon \sim \delta$, so we use $\delta$ in both contexts. Denoting the regularized angular collision integral by $St_{\theta}\left(w,f_{w},\theta_{k},\delta\right)$, we are looking  for a $f_{w_0}$ such that 
 \begin{equation}\label{eq:cauchy}
\lim_{\delta\rightarrow0}St_{\theta}\left(w_0,f_{w_0},\theta_{k},\delta\right)=0.
 \end{equation}
Now, the \textit{Ansatz} \eqref{eq:turbspec} for the energy spectrum makes the main term in the collision integral \eqref{eq:kin sim} equal to %proportional to 
\begin{align}\label{eq:family}
\omega_{\beta}e_{\alpha}e_{\gamma}+\omega_{\gamma}e_{\beta}e_{\alpha}+\omega_{\alpha}e_{\beta}e_{\gamma}=\addMSC{\frac{\sum\limits_{\chi=\alpha,\beta,\gamma}\sigma_{\chi}K_{\chi}^{w}\left(\cos\theta_{\chi}\right)^{1-2f_{w}}}{e_\alpha e_\beta e_\gamma}}
\end{align}
\begin{comment}
\begin{align}\label{eq:family}
\omega_{\beta}e_{\alpha}e_{\gamma}+\omega_{\gamma}e_{\beta}e_{\alpha}+\omega_{\alpha}e_{\beta}e_{\gamma}\propto \frac{\!\!\!\!\!\sum_{\chi=\alpha,\beta,\gamma}\!\!\!\!\!\sigma_{\chi}K_{\chi}^{w}\left(\cos\theta_{\chi}\right)^{1-2f_{w}}}{e_\alpha e_\beta e_\gamma}
\end{align}
\end{comment}
%\addOB{times the totally symmetric $e_\alpha e_\beta e_\gamma$}.
The energy equilibrium solution $(w,f_w)=(0,0)$ makes the main term identically zero because \eqref{eq:family} is then the resonance condition $\sum_\chi\omega_\chi=0$.
Searching for simple power law structures in the collision integral suggests that the curve $1-2f_w=w$ is special, \addMSC{which motivates the inspired guess $(w_0,f_{w_0})=(3,-1)$.  Now $f_{w_0}$ is a constant independent of the angle, suggesting that the turbulent solution exhibits self-similarity in scale and also in frequency. We proceed by computing the regularized collision integral $St_\theta(3,f_w,\theta_k,\delta)$ as follows: First, we parameterize analytically the resonant manifold \eqref{eq:resonant manifold}, reducing the collision integral to a one-dimensional integral (see supplementary material for details). Next, for each fixed $\delta$ and at each value of $\theta_k$, we seek a constant $f_w$ on the real line %within the interval $(-1-\epsilon, -1+\epsilon)$ with some $\epsilon > 0$ 
that zeros the collision integral. The one-dimensional integral is computed numerically (see Mathematica notebook in the supplementary material), and we repeat this process for progressively smaller values of $\delta$. For each $\delta$, we find a unique $f_{w_0}(\delta)$ s.t $St_\theta(3, f_{w_0}(\delta), \theta_k, \delta) = 0$. As $\delta\rightarrow 0$ we see $f_{w_0}(\delta)\rightarrow -1$. The results confirm that $\lim_{\delta \to 0} St_\theta(3, -1, \theta_k, \delta) = 0$, as illustrated in Figure 2. This approach effectively generalizes the concept of the principal value of the collision integral around divergences caused by the vanishing of frequencies. These divergences cancel out, yielding what we refer to as a generalized solution.}

\begin{figure}[h]
\includegraphics[scale=0.42]{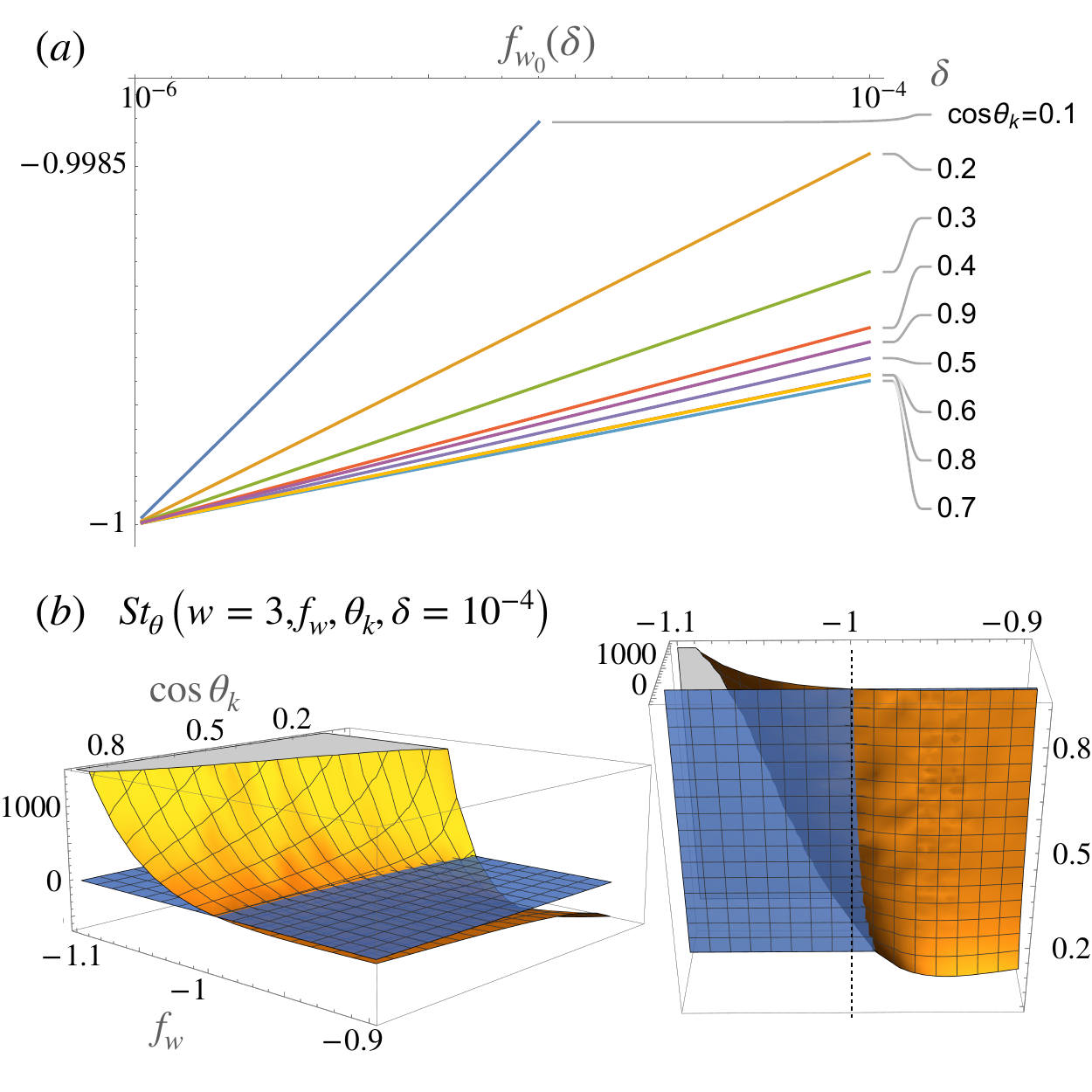} 
\caption{\label{fig:b_delta} Top: Convergence of $f_{w_0}(\delta)\to-1$ as a root of the angular collision integral $St_\theta(w_0=3,f_{w_0},\theta_k,\delta)$ for various angles. Bottom: The collision integral (yellow) at $w_0=3$ and $\delta=10^{-4}$ normalized by its value at $f_{w_{0}}=-1,\cos \theta_k=0.9$ from side and top view  intersecting the zero plane (blue).}
\end{figure}

\textbf{\textit{The turbulent energy spectrum}} %and its flux.}} 
Our result for the  2D wave energy spectrum over $(k_x,k_z)$ in the inertial range between large-scale forcing and small-scale dissipation and assuming zero net pseudo-momentum is 
\begin{equation}
\label{eq:specb2}
e_{\left(\sigma,\mathbf{k}\right)}  = C_0 K^{-3}(\cos\theta_k)^{-2} = C_0 K^{-1} k_x^{-2}.
\end{equation}
It is shown in the supplementary material that for a finite total energy flux $\Pi_0$ the Kolmogorov constant $C_0$ is given by
\begin{equation}\label{eq:KC}
C_0\addMS{\propto \sqrt{\left|\Pi_{0}\right|\delta}} ,
% C_0=\sqrt{\Pi_0/2\pi} log(\epsilon). 
\end{equation}
so as $\delta\to0$ a finite flux $\Pi_0$ can be maintained with a smaller and smaller energy spectrum.
Conversely, at fixed $C_0$ the flux diverges as \addMS{$\Pi_0\sim \delta ^{-1}$}, which is a neat indication for the breakdown of the kinetic description given by \eqref{eq:kin sim} for interactions with and among slow modes. %Sets a bound on the range of applicability of our kinetic description. 
The spectrum is plotted on the regularized domain in Figure 1\ref{fig:2dspec} as a density over $(k_x,k_z)$.  To compare it with the oceanic GM spectrum \cite{munk1981internal} and other theoretical works we can use the full, non-hydrostatic dispersion relation \eqref{eq:omega} to rewrite \eqref{eq:specb2} as a spectral density over positive $(\omega, k_z)$.  This yields  
\begin{align}
C_0 K^{-1} k_x^{-2} \,\mathrm{d}k_x \mathrm{d}k_z &= C_0 N \omega^{-2} k_z^{-2}\,\mathrm{d}\omega \mathrm{d}k_z.
\end{align}
This agrees with the GM spectrum for frequencies large compared to the Coriolis frequency and for vertical wavelengths small compared to the ocean depth. To the best of our knowledge, this is the first time that this limit of the 2D GM spectrum has been derived from first principles; it is certainly the first time non-hydrostatic waves are included in the analysis.     In their empirical model, Garrett and Munk assumed statistical independence between $k_z$ and $\omega$ for simplicity, but there is also observational support for this assumption \cite{dematteis2024interacting}.  \addOBB{Now, in our work this statistical independence is actually strictly implied by the homogeneous \textit{Ansatz} for the $(k_x,k_z)$-spectrum
$e_\alpha = C_0K^{-3}e_\theta(\theta)$, even if the function $e_\theta(\theta)$ is arbitrary.  This follows because $A\mathrm{d}k_x\mathrm{d}k_z = B\mathrm{d}\omega\mathrm{d}k_z$ precisely if $B=AK^3k_z^{-2}/N$, so the corresponding  $(\omega,k_z)$-spectrum is in the separable form $C_0e_\theta(\theta) k_z^{-2}/N$, i.e., the  power law $k_z^{-2}$ multiplied by a function of angle and therefore of internal wave frequency.}

\addOB{\textbf{\textit{Comparison with previous hydrostatic results}} The hydrostatic approximation $k_x^2 \ll k_z^2$ modifies the dispersion relation from $\omega^2=N^2k_x^2/(k_x^2+k_z^2)$ to $\omega^2=N^2k_x^2/k_z^2$.  This is accurate for low-frequency internal waves, with $\omega^2\ll N^2$, but not otherwise.  Indeed, if used outside its range of validity, the modified dispersion relation produces unbounded frequencies as $k_z\to0$, in stark contrast to the exact bound $\omega^2\leq N^2$.   This may lead to unphysical divergences of the collision integral near the $k_x$-axis.  Also, the hydrostatic approximation does not remove the fundamental difficulties of kinetic wave theory with zero-frequency wave modes.     
But the advantage of this approximation is that the frequency and interaction coefficients are now bi-homogeneous functions of $k_x$ and $k_z$ and hence one can look for bi-homogeneous solutions $e_\alpha=C_0  k_x^{2w_x} k_z ^{2w_z}$.  One pair of power-laws $(2w_x,2w_z)$ can be derived using the classic Zakharov—Kuznetsov~(ZK) transformation \cite{pelinovsky1978wave, ZLF}, with $2w_{x}=-w_{\Gamma_{x}}-1-\frac{w_{\omega_{x}}}{2}$ and $2w_{z}=-w_{\Gamma_{z}}-1-\frac{w_{\omega_{z}}}{2}$. Here  $w_{\Gamma_{x/z}}$ and $w_{\omega_{x/z}}$ are the homogeneous degrees with respect to $k_x/k_z$ of the interaction coefficient and frequency. This suggests in our case, with $(w_{\Gamma_{x}},w_{\Gamma_{z}}, w_{\omega_{x}}, w_{\omega_{z}})=(0,1,1,-1)$, the energy spectrum $e_\alpha \sim k_{x}^{-3/2}k_{z}^{-3/2}$ as was derived already 30 years ago for horizontally unidirectional 2D integral gravity waves \cite{daubner1996stationary} and later also for 3D horizontally isotropic waves, with the horizontal wavenumber magnitude $k_\perp$ replacing $k_x$ \cite{caillol_kinetic_2000} in the 2D spectrum.}     \addOB{A subsequent series of kinetic wave studies  (\cite{lvov2001hamiltonian,lvov2010oceanic} and references therein) considered a general family of horizontally isotropic bi-homogeneous power laws, focusing on the convergence of the collision integral.   Only one choice of this family led to a convergent integral, with a corresponding 2D energy spectrum proportional to $k_\perp^{-1.7}k_z^{-1}$.  } \addOB{The recent study \cite{lanchon2023energy} considered induced diffusion triads \cite{mccomas1977resonant} and derived analytically a horizontally isotropic energy spectrum  with a 2D density proportional to $k_\perp^{-1}k_z^{-2}$.  The corresponding $(\omega, k_z)$ spectrum is then proportional to $\omega^{-1}\, k_z^{-2}$, but in this theory the support of the 2D spectrum in spectral space has a non-trivial shape, so arguably this still leads to a 1D marginal spectrum proportional to $\omega^{-2}$.}

\addOB{Now, our solution \eqref{eq:specb2} has the hydrostatic approximation $e_\alpha \sim k_x^{-2}k_{z}^{-1}$, which disagrees with the results of all prior studies.  To reconcile these opposing facts we suggest that it is essential to use the full dispersion relation to obtain results that are uniformly valid in spectral space.  Making the hydrostatic approximation \textit{a priori} implies restricting to low-frequency waves with $k_x^2\ll k_z^2$, but the methods of kinetic wave theory are necessarily global in nature and don't automatically respect this restriction, whether they use conformal mappings or exploit the homogeneity and scaling of the collision integral.   This connects with the aforementioned unphysical divergence of the hydrostatic approximation near the $k_x$-axis.   
In summary, the facts suggests that one needs to consider the full dispersion relation in order to find the full turbulent solution \eqref{eq:specb2} or even its hydrostatic limit.}

\textit{\textbf{Conclusion and perspectives.}}
\addOB{Our work shows how a divergence of the collision integral on the set of vanishing frequency, which is a generic feature of many anisotropic dispersive equations, can be overcome by regularizing the collision integral and then studying the limit of vanishing regularization.  This is a new method with potential applicability to other physical systems such as waves in plasma and also 3D internal gravity waves.  In the present case of 2D internal gravity waves it produced the unique turbulent energy spectrum $e_{\left(\sigma,\mathbf{k}\right)} =C_0 K^{-3}(N/\omega_k)^{2}$.}   This spectrum holds in the inertial range between large-scale forcing and small-scale dissipation provided the forcing does not produce net pseudo-momentum on average; if the latter is not true then one can expect subtle modifications of the results \cite{shavit2024sign}, which is a topic for future research.  For vertical scales short compared to the depth of the ocean and frequencies large compared to the Coriolis frequency our 2D spectrum agrees with the 2D oceanic GM spectrum.  Apparently, this is the first time that a theoretical internal wave spectrum agrees with the \addMS{2D} GM spectrum.   Of course, actual ocean observations exhibit a range of power laws somewhat close to the GM spectrum, so this should not be overstated \cite{lvov2010oceanic,dematteis2024interacting}.  \addOB{We have experimented numerically with generalizations of \eqref{eq:turbspec} a finite distance away from $w=3$ and found good evidence that a one-parameter continuous family of steady states exist for values of $2<w\leq 3$ combined with the choice $2f_w=1-w$.   This family includes the convergent steady state identified in \cite{lvov2010oceanic}  at $w=2.7$.  However, as with all power law steady states away from $w=3$, there is no energy flux that we can associate with that steady state.}
A natural progression would add Coriolis forces to the equations, which would serve as a physical regulator by decoupling slow modes from waves due to the gap in the rotating dispersion relation, $\omega^2=N^2\cos^{2}\theta+f^{2}\sin^{2}\theta$, where $f$ is the Coriolis parameter. \addMSS{Adding rotation does not change the homogeneity degree of the interaction coefficients and frequency, hence the homogeneous part of the turbulent spectrum is expected to remain whilst the angular part might change: $e_\alpha=C_0 K^{-3}e_\alpha^{\theta}(\omega,f/N)$. The new parametric dependence on $f/N$ might play an important role in the variability of the observed ocean spectra.}

\textit{\textbf{Acknowledgments}}.
We thank Gregory Falkovich \addOB{and Pierre-Philippe Cortet} for useful discussions. Special thanks are owed to Vincent Labarre for his invaluable insights and extensive feedback, which significantly contributed to the development of this manuscript. 
\addOB{The referees' comments significantly improved the original manuscript.} 
This work was supported by the Simons Foundation and the Simons Collaboration on Wave Turbulence.
  %The numerical study was made
  %possible thanks to New York University's Greene computing
  %cluster facility.
  OB acknowledges additional financial support under ONR grant N00014-19-1-2407 and NSF grant DMS-2108225. MS acknowledges additional financial support from the Schmidt Futures Foundation. %and the Israeli CHE. 

\bibliography{main}% Produces the bibliography via BibTeX.

\newpage

\begin{comment}

\appendix
\section{The collision integral}
\label{appendix}

\textit{\textbf{The collision integral}}.

The momentum conservation in the vertical direction splits the angular integration into two parts $St_\theta=St_\theta^{I}+St_\theta^{II}$, where in $I$: $\sin\theta_\beta<0$ and $\sin\theta_\gamma>0$ and in $II$: $\sin\theta_\beta<0$ and $\sin\theta_\gamma<0$. 
The appendix includes the parametrization of the resonant manifold and the explicit form of the collision integral. We find that for small $\delta$ there is an isolated value $b_{I}(\delta,\theta_k)$ for each $\theta_k$ s.t $St_\theta^I(b_I,\theta_k)=0$ and the same for $St_\theta^{II}$. We find that $b_{I}(\delta,\theta_k)$ is approaching $-1$ from the right, while $b_{II}(\delta,\theta_k)$ is approaching zero from the left. \addMS{this needs to be moved to the appendix}

\end{comment}

\end{document}